\documentclass[twocolumn,prl,showpacs]{revtex4}
\usepackage{amsmath}
\usepackage{amssymb}
\usepackage{graphicx}

\begin{document}

\title{Intrinsic Cavity QED and Emergent Quasi-Normal Modes for Single Photon}

\author{H. Dong$^{1}$, Z. R. Gong$^{1}$, H. Ian$^{1}$, Lan Zhou$^{2,3}$, and C. P. Sun$^{1}$}

\affiliation{$^{1}$Institute of Theoretical Physics, Chinese Academy of Sciences,
Beijing, 100080, China}

\affiliation{$^{2}$Frontier Research System, The Institute of Physical and Chemical
Research (RIKEN), Wako-shi 351-0198, Japan}

\affiliation{$^{3}$Department of Physics, Hunan Normal University, Changsha 410081,
China}

\begin{abstract}
We propose a special cavity design that is constructed by terminating
a one-dimensional waveguide with a perfect mirror at one end and doping
a two-level atom at the other. We show that this atom plays the intrinsic
role of a semi-transparent mirror for single photon transports such
that quasi-normal modes (QNM's) emerge spontaneously in the cavity
system. This atomic mirror has its reflection coefficient tunable
through its level spacing and its coupling to the cavity field, for
which the cavity system can be regarded as a two-end resonator with
a continuously tunable leakage. The overall investigation predicts
the existence of quasi-bound states in the waveguide continuum. Solid
state implementations based on a dc-SQUID circuit and a defected line
resonator embedded in a photonic crystal are illustrated to show the
experimental accessibility of the generic model.
\end{abstract}


\maketitle

\textit{Introduction}-- Cavity quantum electromagnetic (QED)
\cite{berman} essentially reveals the quantum nature of light photon
confined in an extremely small spatial volume, such that there
exists a strong coupling between atom and electromagnetic(EM) field.
The atomic spontaneous emission within a micro-cavity can be
enhanced or suppressed. The very property reflects the coherent
manipulation of atom - EM field interactions through the boundary
condition of cavity. When the cavity boundary is leaky, the set of
quasi-normal modes (QNM's), which is introduced in the study of the
scattering of gravitational waves by a Schwarzschild black
hole~\cite{Vishveshwara}, appears as a discrete spectrum with
complex value. Its imaginary part represents the width of the
resonant spectrum line~\cite{scully,Young}.

In a usual experimental setup, the cavity is bounded by two
reflective mirrors, which are obviously the external objects
independent of the atom inside. In this letter, we consider an
alternative cavity approach by localizing an atom at a position
inside a one-dimensional (1D) half-resonator waveguide. The cavity
is bounded by the termination of the waveguide at one end and the
localized atom at the other. Unlike a few two- or three-dimensional
half-cavity setups~\cite{semi,zoller}, the characteristic intrinsity
of this two-end cavity design arises naturally, in which the atom
provides a tunable boundary condition for the cavity EM field,
controllable through the parameters of the atom. When the atom
resonates with the cavity field, the interference effect between its
emission and absorption of photons tunes the boundary to totally
reflect the incident EM wave. Hence the atom serves as a perfect
mirror and the normal mode emerges. Near resonance, the atom behaves
like a semi-transparent mirror, resulting in a leaky cavity and the
emergence of QNM's. In either case, there exists a strong
back-action from the emerging cavity field on the atom whose
spontaneous emission is thus evidently influenced.

The theoretical considerations above inspire us to design a quantum
coherent device, which has the atom responsible for a quantum
switch~\cite{liuyx} and stores a single photon as the QNM's or the
normal modes of the tunable cavity. The recently proposed single
photon transistor~\cite{lukin07} was designed upon a similar footing
by utilizing the surface plasmon excitation confined in an infinite
waveguide but without any mirror~\cite{lukin-prl}.

We further explore two alternative implementations of more
laboratory accessibility of the original design, one of which uses a
superconductive transmission line resonator and the other is based
on a defected line resonator within a photonic crystal. These
implementations are useful for showing the detailed phenomena
related to the QNM's theory~\cite{scully,Young}.

\begin{figure}
\includegraphics[bb=24 604 257 758, width=7cm]{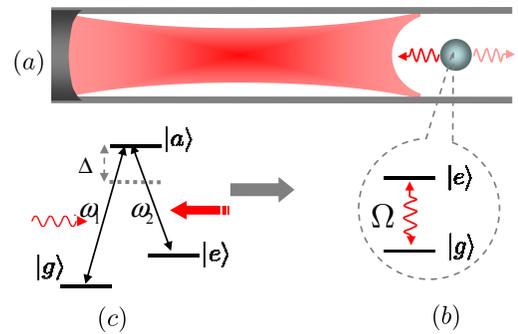}
\caption{(\textit{Color online}) Schematic of the intrinsic cavity.
Half-waveguide(a) with a two-level atom as a tunale mirror(b); To
implement the steady two-level atom, we use the stimulated Ramma
process based on a $\Lambda$-type atom(c) behaves as a two-level
atom in large detuning to overcome the high-level decay.}
\label{fig:modelsetup}
\end{figure}

\textit{The Generic Model} -- We exhibit our idea about the
intrinsic cavity by way of example shown in
Fig.~\ref{fig:modelsetup}. The system consists of a 1D semi-infinite
waveguide (a half-cavity) with a two-level atom localized at a
distance $a$ from the termination. The atom has a level spacing
$\Omega$ between its ground state $\left\vert g\right\rangle $ and
its excited state $\left\vert e\right\rangle $ and is coupled to an
EM field with strength $J$. Similar to the case of a finite or
two-end cavity in Ref.~\cite{Fan}, the model Hamiltonian of our
setup reads

\begin{eqnarray}
H & = & -iv_{g}\int_{0}^{\infty}\mathrm{d}x(\varphi_{R}^{\dagger}\partial_{x}\varphi_{R}-\varphi_{L}^{\dagger}\partial_{x}\varphi_{L})+\Omega\left\vert e\right\rangle \left\langle e\right\vert \nonumber \\
 &  & +J[\varphi_{R}^{\dagger}(a)+\varphi_{L}^{\dagger}(a)]\left\vert g\right\rangle \left\langle e\right\vert +h.c.\end{eqnarray}
 where $\varphi_{R}=\varphi_{R}(x)$ ($\varphi_{L}=\varphi_{L}(x)$)
is the bosonic field operator for a right-going (left-going) photon;
$v_{g}$ stands for the group velocity of the photon in the waveguide.

The state of the system with either a single photon or an excited
atom spans an invariant subspace of $H$. Thus, corresponding to the
eigenvalue $E=ck$, the stationary eigenstate

\begin{equation}
\left\vert E\right\rangle =\int_{0}^{\infty}\mathrm{d}x\left(u_{k,R}\varphi_{R}^{\dagger}+u_{k,L}\varphi_{L}^{\dagger}\right)\left\vert 0,g\right\rangle +w_{k}\left\vert 0,e\right\rangle ,\end{equation}
depicts a single photon transport in the half-cavity, where $\left\vert 0,g\right\rangle $
indicates the state without photon while the atom stays grounded and
$\left\vert 0,e\right\rangle $ the state without photon while the
atom is excited. For a photon of momentum $k$, the probability amplitudes
$u_{k,R}=u_{k,R}(x)$, $u_{k,L}=u_{k,R}(x)$, and $w_{k}$, corresponding
respectively to the right-going photon, the left-going photon, and
the excited atom, satisfy the following equations:

\begin{eqnarray}
-iv_{g}\partial_{x}u_{k,R}+J\delta\left(x-a\right)w_{k} & = & Eu_{k,R},\\
iv_{g}\partial_{x}u_{k,L}+J\delta\left(x-a\right)w_{k} & = & Eu_{k,L},\\
J\left[u_{k,R}\left(a\right)+u_{k,L}\left(a\right)\right] & = & \left(E-\Omega\right)w_{k}.\end{eqnarray}

\begin{figure}
\includegraphics[bb=24 632 355 775, width=8cm]{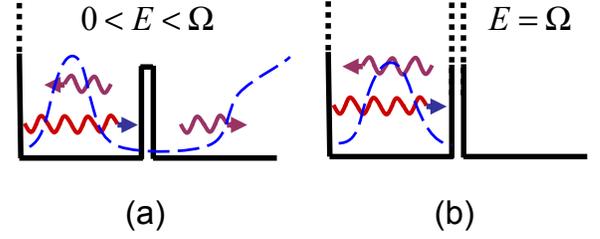}
\caption{(\textit{Color online}) The emergence of the QNM's. The
\textit{wavy line} represents the reflective properties of the atom;
the \textit{dashed line} shows the typical form of the photon wave
function at resonance. (a) when $0<E<\Omega$, the model is
equivalent to a leaky cavity. (b) when $E=\Omega$ and
$Ea/v_{g}=n\pi$, the area is well confined and the cavity completely
reflects the photon.} \label{fig:potentials}
\end{figure}

To understand the meaning of these amplitudes, we apply the sum of
the amplitudes
$\varphi_{k}\left(x\right)=u_{k,R}\left(x\right)+u_{k,L}\left(x\right)$
of the EM field to the Maxwell equation~\cite{Young}:
\begin{equation}
\partial_{x}^{2}\varphi_{k}(x)=-\left(\frac{E}{v_{g}}\right)^{2}\rho(x)\varphi_{k}(x),\label{eq:emmove}\end{equation}
 where \begin{equation}
\rho\left(x,E\right)\equiv1+\frac{2J^{2}\delta\left(x-a\right)}{E\left(\Omega-E\right)}\end{equation}
is the singular refractive index over the waveguide, which depends
on the energy of the system. This energy dependence is illustrated
in Fig.~\ref{fig:potentials}. Eq.~(\ref{eq:emmove}) can be regarded
as a schr\"{o}dinger equation where
$V\left(x,E\right)\equiv\rho\left(x,E\right)-1$ plays the role of an
energy-dependent local potential. The resonance effect occurs at
$E=\Omega$ where a $\delta$-type infinite effective potential will
emerge. Thus, the atom and the fixed mirror at the termination
sandwich a confined 1D space to form a perfect cavity when the
cavity photon mode of energy $E$ resonates with the level jumps
$\Omega$ of the atom. In other words, the resonant atom and cavity
gives rise to a spectrum of normal modes. In the case of near
resonance, the atom plays the role of a leaky mirror whose rate of
leakage is determined by refractive index $\rho(x,E)$ which in turn
is controlled by the parameters of the atom (such as the level
spacing and the coupling constant to the EM field). Actually, the
emergence of QNM's in a leaky cavity has been widely investigated
for a given refractive index~\cite{Young}, but the alternative we
offer here in a 1D continuum bears the spontaneous emergence of
QNM's where the effective $\delta$-type potential is given by the
localized atom.

\begin{figure}
\includegraphics[clip,width=8.5cm]{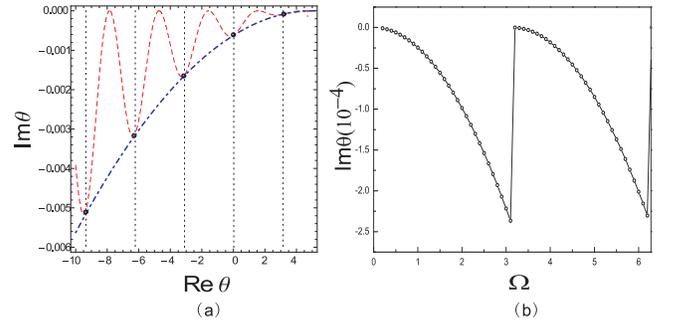}
\caption{(\textit{Color online}) (a) The solution of
Eq.\ref{eq:simpified}. The roots of Eq. \ref{eq:simpified} are
marked by the circle. The analytical solution under approximation is
shown by a dot-dash line (blue). The approximation fits numerical
point. (b) $\textbf{Im}\theta$ vs atom frequency. The decay rate
reaches its minimum, when $\Omega=n\pi$.} \label{fig:sol-reson}
\end{figure}

\textit{Quasi-normal Modes as Quasi-bound State} -- Following the
routine treatment of QNM's, we assume the field amplitude outside
the cavity ($x>a$) as an outgoing plane wave
$\varphi_{k}\left(x\right)=B\exp\left(ikx\right)$ and perform an
analytical continuation by taking the photonic momentum $k=E/v_{g}$
as a complex number~\cite{Young}. The outgoing plane wave here
describes the spontaneous radiation of the atom. The corresponding
field amplitude $\varphi_{k}\left(x\right)=A\sin\left(kx\right)$
inside the emergent cavity ($0<x<a$) is identical to that of a
perfect cavity except that $k$ is complex. Accordingly, the energy
of the QNM's is a complex quantity with its real part being the
resonant frequency and its imaginary part the linewidth.

This linewidth determines the leakage rate of the EM field from the
cavity~\cite{nakamura07}. The boundary conditions for the field
amplitude and its first derivative give the dimensionless equation

\begin{equation}
\tan\left(\theta\right)=\left(\frac{\kappa}{W-\theta}+i\right)^{-1}\label{eq:simpified}\end{equation}
which determines the complex energy of the QNM's. Here, $\theta=Ea/v_{g}$,
$W=\Omega a/v_{g}$ and $\kappa=2J^{2}a/v_{g}^{2}$ denotes a set
of dimensionless parameters. To confine the EM field so that an equivalent
cavity emerges, the interaction between the EM field and the atom
should be sufficiently large, i.e. $1/\kappa\ll1$. Then the QNM's
have their energies approximated by the dimensionless complex eigenvalues
$E_{j}\simeq\omega_{j}-i\gamma_{j}$ where, for $j=0,\pm1,\pm2,....$,

\begin{equation}
\omega_{j}=\frac{v_{g}}{a}\left[j\pi+\frac{1}{\kappa^{2}}\left(W-j\pi\right)\left(\kappa-1\right)\right]\end{equation}
defines the resonant frequencies while \begin{equation}
\gamma_{j}=-\frac{v_{g}}{a\kappa^{2}}\left(W-j\pi\right)^{2}\end{equation}
are the linewidths of the modes.

The discreteness of the real part $\omega_{j}$ is similar to the
case of a perfect cavity with normal modes $\omega_{j}^{0}=jv_{g}\pi/a$.
In both cases, the cavity length $a$ determines the discrete values
of the modes. This property results exactly from the so-called shape
resonance: when the length $a$ matches the frequency of the EM wave
inside the cavity, the reflective wave from the ``atom boundary''
contains a phase retardation of $\pi$ relative to the incident wave,
destructively interfering the incident wave.

The negativity of the imaginary part $\gamma_{j}$ characterizes the
decaying of the EM field from inside cavity to the outside. We define
the lifetime of the EM field as the inverse $\tau=|\mathbf{\mathrm{Im}}E_{j}|^{-1}$,
i.e. \begin{equation}
\tau=\left(\frac{\kappa}{W-j\pi}\right)^{2}\frac{a}{v_{g}}.\label{eq:decaytime}\end{equation}
The lifetime is proportional to $J^{4}$ and the inverse square of
the detuning, which reaches its maximum at $j=\left\lfloor W/\pi\right\rfloor $.
This long-life state is selected by the atomic level spacing $\Omega$.
The effective refraction index $\rho\left(x\right)$ is strongly modified
when the system is close to resonance and the confinement of the EM
field seems much tighter. In this sense, the near-resonant photonic
mode is reflected more fiercely back to the cavity and the leakage
is suppressed. Therefore, it is expected that, when the atomic level
spacing $\Omega$ matches the cavity length $a$ with the value $\Omega=k\pi v_{g}/a$
($k\in\mathbb{Z}$), the atom will act like a perfect mirror.

\textit{Numerical Results and Suppressed Emission} -- Furthermore,
to verify the approximated solution of Eq.~(\ref{eq:simpified})
above, we solve it numerically and demonstrate the real part and the
imaginary part of the solution in the $\theta-$complex plane in
Fig.~\ref{fig:sol-reson}(a). The circle points is solved as the
cross points of two lines, red dashed line and black dotted line,
for the parameters $\kappa=200$ and $W=5$. The analytic solution
approximately obtained above is plotted as the dotted dashed blue
line, which fits very well the the numerical solution in the
Fig.~\ref{fig:sol-reson}(a).To show this matching effect in details,
we plot the curve of the decay rate as a function of the atomic
level spacing in Fig.~\ref{fig:sol-reson}(b).

\begin{figure}
\includegraphics[width=5cm]{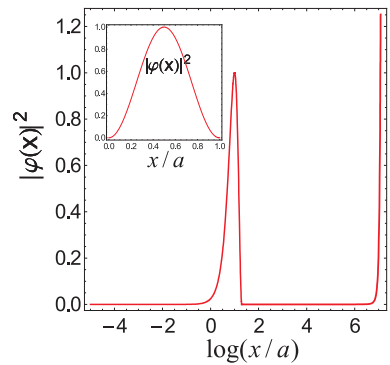}
\caption{(\textit{Color online}) Plot of the wave function of the
slowest decay state in real space. The amplitude of the wave
function diverges in the infinite far away. The QNM state here is
actually a resonant state.} \label{fig:wave}
\end{figure}

To further describe the QNM's obtained above, we plot the wave
function with parameters: $\kappa=200$ and $W=5$ in
Fig.~\ref{fig:wave}, which describes the slowest decay process for
definite $\Omega$. Here, the resonant frequency
$\omega_{0}=3.15084v_{g}/a$ with the decay rate
$\gamma_{0}=-8.5\times10^{-5}v_{g}/a$. There appears a bound state
localized in the intrinsic cavity, however, which diverges
exponentially far away outside the cavity. Those characters
demonstrate that QNMs are actually resonant states~\cite{hatano07}.
Due to the long lifetime, it behaves like a bound state even in a
large time scales. With these consideration, we conclude that there
actually exists a \textit{quasi-bound
state}~\cite{nakamura07,hatano07}. Interestingly, if a single photon
is loaded into this cavity, it would takes a long time to escape
from the cavity. Thus we can propose this artificial fabrication as
a storage device for single photon.

To display the quantum nature of the emergent cavity, we study the
prominent cavity QED effect -- the modification of spontaneous
emission rates within the cavity. The radiation of the bounding atom
can be controlled by the boundary conditions of the EM field. To
this end, we introduce a phenomenological emission rate $\Gamma$ to
the atom, which characterizes the system decay into all other
channels of the environment. Then by replacing $\Omega$ with
$\Omega-i\Gamma$, we formally obtain the decay rate of the atom in
the leaky cavity \begin{equation}
\Gamma_{t}\simeq|\frac{W-j\pi}{\kappa}|^{2}+\frac{\Gamma}{\kappa}-\frac{\Gamma}{\kappa^{2}}.\end{equation}
In the so-called ``strong coupling regime'' where $\kappa\gg1$, the
spontaneous decay is explicitly suppressed. A similar result has
been obtained in a 3D setup~\cite{zoller}, when the atom is placed
on a node of the EM field.

\begin{figure}
\includegraphics[width=8cm]{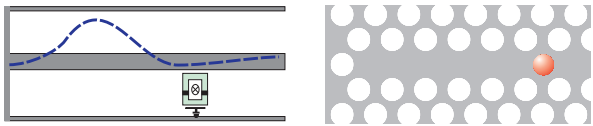}
\caption{(\textit{Color Online})The implementations using,
respectively, a superconducting transmission line resonator and a
photonic crystal.} \label{fig:scoc}
\end{figure}

\textit{Physical Implementations} -- The emergence of QNM's in the
half-cavity predicted above can be materialized through some practical
laboratory systems. Two such examples, shown in Fig.~\ref{fig:scoc},
are the superconducting transmission line resonator with a dc-SQUID-based
charge qubit and a photonic crystal with a doped $\Lambda$-type three-level
atom, in which charge qubit and the three-level atom, respectively,
take the roles of the functional mirror.

In the former, the half-cavity can be realized through the
semi-infinite superconducting transmission line while the two-level
system through the charge qubit with energy eigenstates $\left\vert
a\right\rangle $ and $\left\vert g\right\rangle $. The energy level
spacing $\Omega=\sqrt{B_{z}^{2}+B_{x}^{2}}$ is defined by the field
intensities $B_{z}=4E_{c}(2n_{g}-1)$ and
$B_{x}=2E_{J}\cos(\pi\Phi_{x}/\Phi_{0})$ where
$E_{c}=e^{2}/\left[2(C_{g}+2C_{J})\right]$ is the charge energy and
$n_{g}=C_{g}V_{g}/2e$ is the number of charges at the gate. Note
that $B_{z}$ can be controlled by the voltage $V_{g}$ applied to the
gate capacitance $C_{g}$ whereas $B_{x}$ can be controlled by the
external magnetic flux $\Phi_{x}$ through the SQUID loop. If the
effective length of the transmission line resonator is $L$, the
coupling strength between the qubit and the resonator reads
$V=e\sin\theta C_{g}/C_{\Sigma}\sqrt{\omega/\left(Lc\right)}$. Here,
$c$ is the capacitance per unit length of the transmission line and
$\omega$ is the frequency of the quantized EM field. The coupling
strength ranges from 5 to 200 $MHz$ and the qubit level spacing is
in the range of 5-15 $GHz$~\cite{wallraf}. So long as the coupling
between the charge qubit and the line resonator becomes strong, the
quasi-bound state will appear in the transmission line due to the
enhancement of reflection by the functional mirror~\cite{zhoulan07}.
The coupling strength should reach about several $GHz$ for the
observation of the long-life quasi-bound state.

As for the implementation on a photonic crystal, the doped
$\Lambda$-type atom in its excited state can decay so fast that we
need resort to the stimulated Raman scattering (SRS) technology.
With a coupling constant $G$ and a detuning $\Delta$, a strong drive
field is applied to couple the metastable state $\left\vert
a\right\rangle $ (with less decay rate) and the excited state
$\left\vert e\right\rangle $ with much larger decay rate. The gourd
state $\left\vert a\right\rangle $ and the excited state $\left\vert
e\right\rangle $ is coupled by the EM field inside cavity with
coupling constant $g$. In the case with large detuning for the SRS
happens, an effective tunable coupling $J=-gG/2\Delta$ is induced
between EM field and the ground state $\left\vert g\right\rangle $
and the metastable state $\left\vert a\right\rangle $ . It can be
observed from Eq.~(\ref{eq:decaytime}) that the decay time for
single photon in the cavity is proportional to $V^{4}$. This high
sensitivity to the coupling strength makes it possible to control
the leakage by solely adjusting the intensity of the drive field.

\textit{Conclusion} -- we conceive the architecture of a quantum
device for coherent manipulation at the single photon level, which
is shown realizable in a few practical systems. Our theoretical
approach predicts the intriguing phenomenon that an atom inside a 1D
half-cavity can act as an end mirror to close the half-cavity. The
emerging two-end cavity induces a back-action to influence the
quantum dynamics of the atom. As a result, there forms naturally a
set of quasi-normal modes in the 1D continuum for the single photon
emission process. In addition, the cavity with the emergent singular
refractive index is equivalent to the extremely inhomogeneous space
equipped with a space-time singularity for wave propagation. We
hence expect that a solid-state based implementation can be employed
as a laboratory simulation of the event horizon and the QNM
radiation of a black hole~\cite{blackhole}.

This work is supported by NSFC No. 90203018, No. 10474104, No. 60433050,
and No. 10704023, NFRPC No. 2006CB921205 and 2005CB724508.

\end{document}